\documentclass[seceq]{ptptex}

\usepackage{graphicx}
\usepackage{wrapft}

\title{
X-ray binaries and ultra-luminous X-ray sources in nearby and distant
galaxies  
}

\author{
Marat  \textsc{Gilfanov}%
}

\inst{
Max-Planck-Institut f\"ur Astrophysik, Garching, Germany\\
Space Research Institute, Moscow, Russia
}

\abst{
We review recent results on populations of compact X-ray sources in
normal galaxies. The luminosity distributions of low
and high mass X-ray binaries in nearby galaxies appear to be described
by the respective ``universal'' luminosity functions, whose shapes do not
vary significantly from galaxy to galaxy and the normalizations  are
 proportional to the SFR (HMXBs) and  
stellar mass (LMXBs) of the host galaxy. There is a significant
qualitative difference between   XLFs of high and low
mass X-ray binaries,  reflecting the difference in the accretion
regimes in these two types of X-ray binaries. 

The high luminosity cut-off is by an order of magnitude larger for
HMXBs -- 
although bright sources, $\log(L_X)\ge 39$, are observed in both young
and old galaxies, the truly ultraluminous ones, with $\log(L_X)\sim
40-40.5$, are associated 
with the regions of intense star formation and have not been detected
so far in old stellar populations of elliptical and S0 galaxies. 
The $L_X-$SFR relation for distant galaxies in the
HDF-N indicates that  ULXs at  redshifts of $z\sim 0.2-1.3$
were not significantly more luminous than those observed in nearby
galaxies. 
}

\begin{document}

\maketitle

\section{Introduction}

Chandra observatory, thanks to its sub-arcsec angular resolution,
opened a new era in studying X-ray binary populations in nearby
galaxies. For the first time an opportunity was presented to 
observe compact sources in a nearly confusion free regime. 
The long suspected fact has been proved, that X-ray binaries are an 
important, if not dominant, contributor to the X-ray emission of the normal
galaxies,\cite{fabbiano2003} as illustrated by the example of our
Galaxy.\cite{grimm02}  

Depending on the mass of the optical companion, X-ray binaries are
subdivided in to two classes -- low  and high mass X-ray 
binaries, having significantly different evolutionary time
scale, $\sim 10^{6-7}$ and $\sim 10^{9-10}$ years respectively 
\cite{xrbrev}. 
The nearly prompt emission of HMXBs  makes them a potentially  good
tracer of the recent star formation activity in the host galaxy
\cite{rs78}.  The LMXBs, on the other hand, have no relation to the 
present star formation, but, rather, are related to the stellar
content of the host galaxy \cite{lmxb}.
Chandra observations of the nearby galaxies presented a possibility to 
confirm this simple picture and to calibrate the
HMXB--SFR\cite{grimm03,lx-sfr,ranalli} 
and LMXB$-{\rm M_*}$\cite{colbert04,lmxb,kim2004} relations.

An unusual class of compact sources -- ultraluminous X-ray sources,
has been discovered in nearby galaxies more than a decade ago
\cite{colbert99,fab89}.  
Although bright, $L_X>10^{39}$ erg/s, point-like sources are found
both in young star forming galaxies and in old stellar population of
elliptical and S0 galaxies, the most luminous and exotic objects are
associated with actively star forming galaxies.  
Their nature and relation to more ordinary X-ray binaries  
is still a matter of a significant debate. Based on a  
simple Eddington luminosity argument, they appear to be powered by
accretion onto an intermediate mass object -- a black hole with the
mass in the hundreds-thousands solar masses
range\cite{fabbiano2003,miller2003}. However, a number 
of alternative models have been considered as well -- from collimated 
radiation\cite{koerding2002} to  $\sim$stellar mass black holes,
representing the high mass tail of the standard stellar evolution
sequence and accreting in the near- or slightly super-Eddington
regime\cite{king2001}.

\begin{figure}
\centerline{\hbox{
\resizebox{0.45\hsize}{!}{\includegraphics{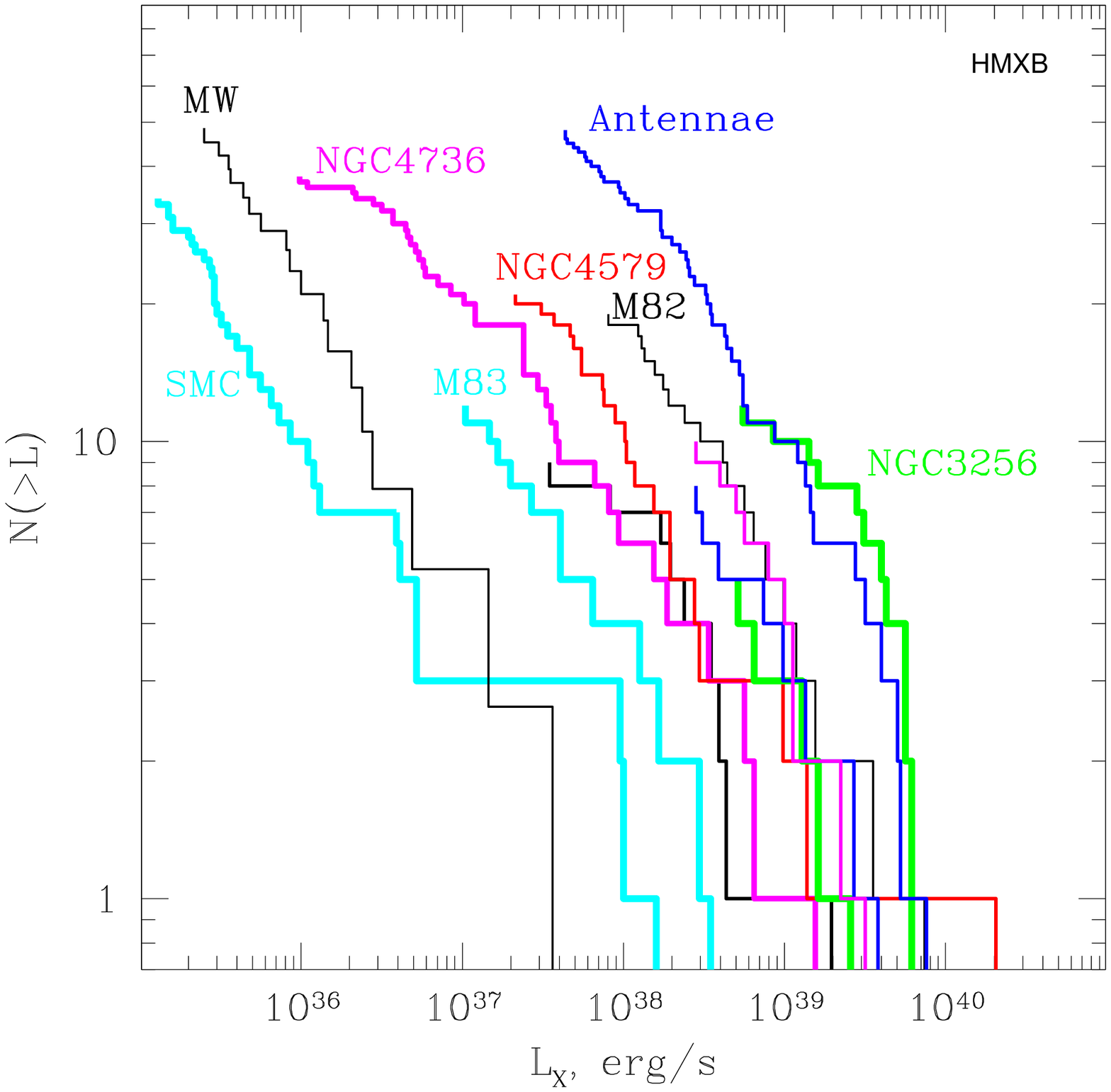}}
\resizebox{0.45\hsize}{!}{\includegraphics{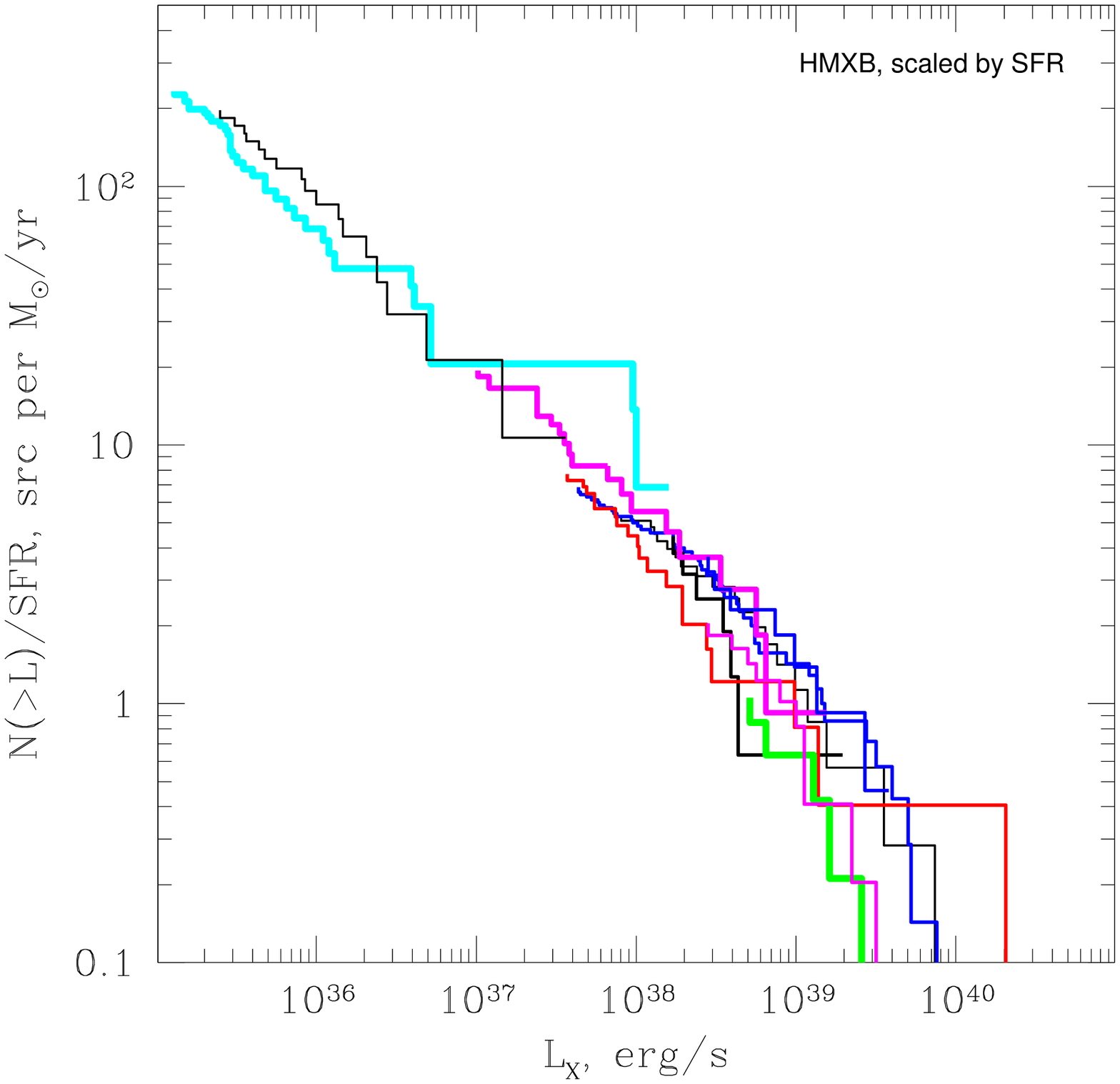}}
}}
\caption{{\em Left:} The XLFs of compact X-ray sources in nearby star
forming galaxies. 
{\em Right:} The XLFs of the same galaxies normalized to the star
formation rates (from Grimm et al.\cite{grimm03}).
The XLF of the Milky Way includes only HMXBs.
}   
\label{fig:lf_hmxb}
\end{figure}

\begin{figure}
\centerline{\hbox{
\resizebox{0.45\hsize}{!}{\includegraphics{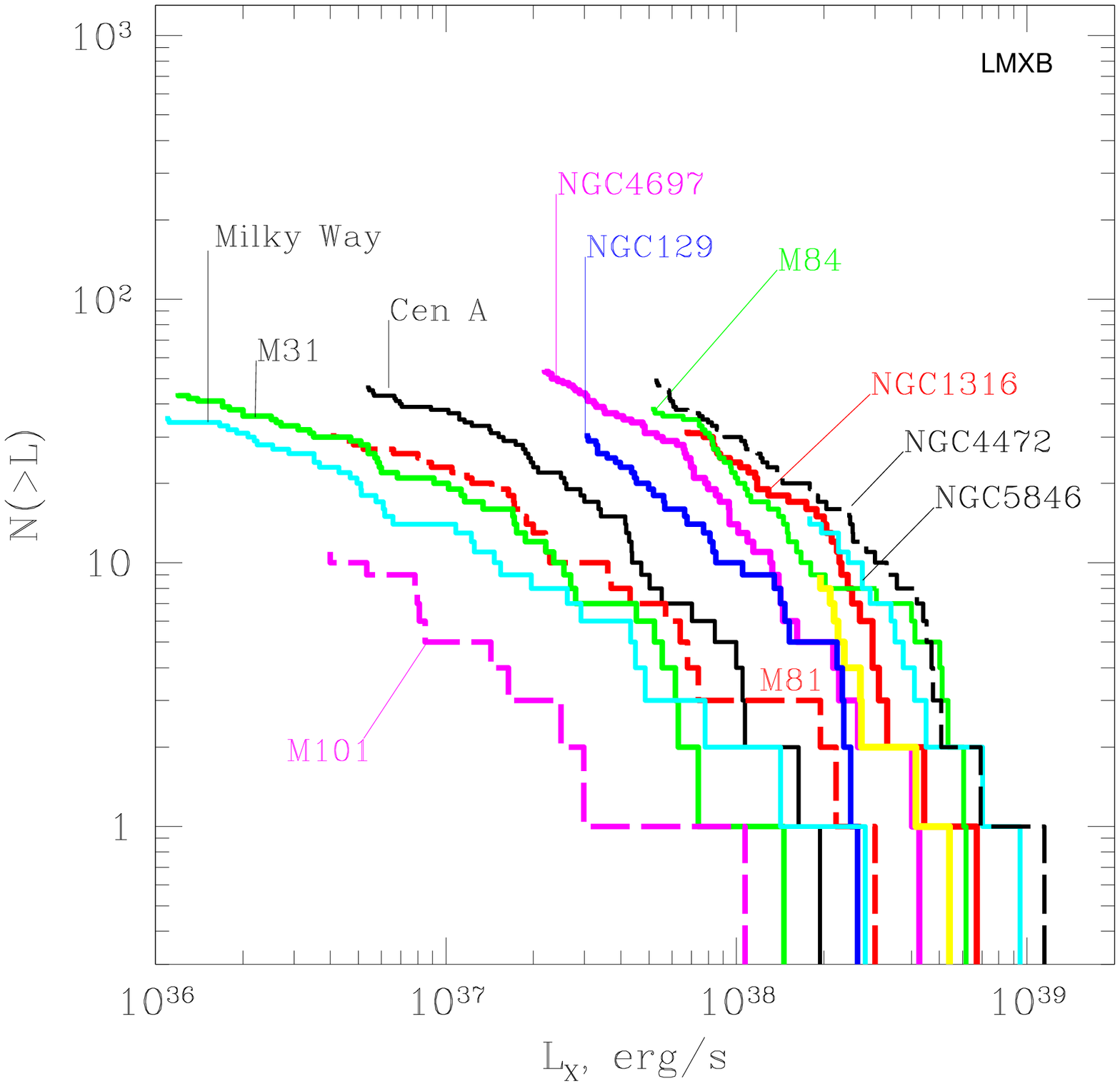}}
\resizebox{0.45\hsize}{!}{\includegraphics{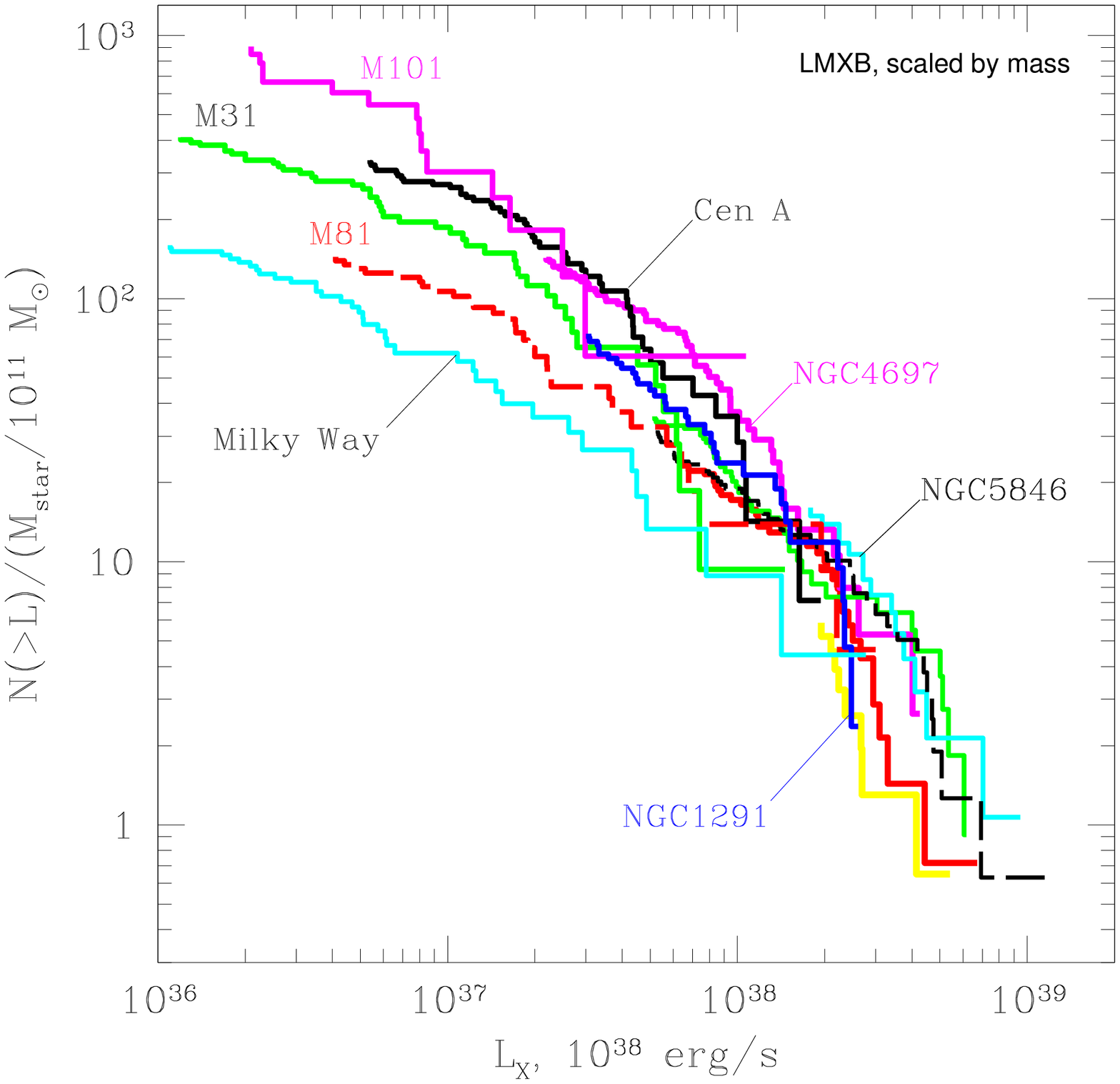}}
}}
\caption{{\em Left:} The XLFs of compact X-ray sources in nearby
elliptical, S0 galaxies and bulges of spiral galaxies.  
{\em Right:} The same XLFs normalized to the stellar mass of the host
galaxy (from Gilfanov\cite{lmxb}). 
The XLF of the Milky Way includes only LMXBs.
}
\label{fig:lf_lmxb}
\end{figure}

\section{``Universal'' XLFs of high- and low- mass X-ray binaries}
\label{sec:xlf}

The Fig.\ref{fig:lf_hmxb} (left panel) shows observed luminosity
functions (XLFs) of compact X-ray sources in nearby star forming
galaxies from the sample of Grimm et al.\cite{grimm02}. 
There is a large spread in the number of sources and in the luminosity
of the brightest source among the galaxies. 
However, normalized to the SFR of the host
galaxy, the XLFs   match each other both in the slope  and in the
normalization (right panel in  
Fig.\ref{fig:lf_hmxb}). Although a finite dispersion might still
remain, especially at the high luminosity end, the normalized luminosity  
functions occupy a rather narrow band in the $N-L_X$ plane, despite of
the large dynamical range of the star formation rates, a factor of 
$\sim 50$. 
Similarly, the XLFs of the compact sources in the old stellar systems
--  elliptical, S0 galaxies and bulges of spiral galaxies differ
significantly from each other, but the 
\begin{wrapfigure}{l}{6.6cm}
\resizebox{7cm}{!}{\includegraphics{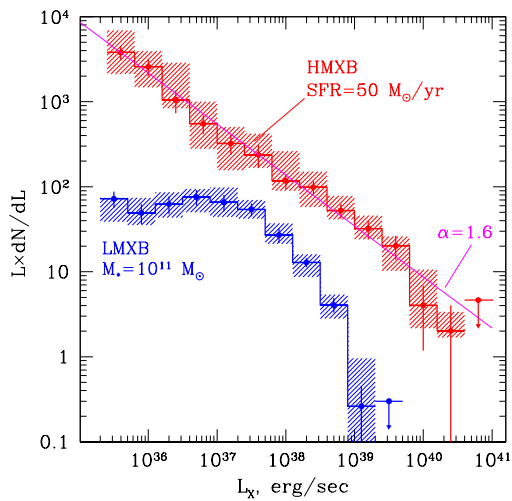}}
\caption{The ``universal'' XLFs of high- and low-mass X-ray binaries
in nearby galaxies (from Grimm et al.\cite{grimm03} and
Gilfanov\cite{lmxb}). The upper limits are at 90\% confidence for a
Poisson distribution.
The shaded areas illustrate the amplitude of systematic errors 
(90\% confidence level) due to uncertainties in the source distance 
(assuming 20\% relative uncertainty), mass-to-light ratios (30\%) and
star formation rates (30\%). }  
\label{fig:xlf_dif}
\end{wrapfigure}
dispersion  decreases notably
after they are normalized to the stellar mass of the
host galaxy (Fig.\ref{fig:lf_lmxb}). 
These examples suggest that the luminosity distributions of HMXBs and
LMXBs are described, to  first approximation, by the ``universal''
XLFs, whose shapes are not subject to significant
galaxy-to-galaxy variations. Their normalizations are proportional,
respectively, to the SFR and stellar 
mass of the host galaxy.\cite{grimm03,lmxb} 

The shapes of the universal XLFs of high- and low-mass X-ray binaries
in nearby galaxies are qualitatively different from each
other (Fig.~\ref{fig:xlf_dif}). The HMXB XLF can be approximated by a 
power law with 
differential slope of $\alpha\approx 1.6$ in a broad luminosity range,
$\log(L_X)\sim 35.5-40.5$ and has a cut-off at $\log(L_X)\sim 40.5$. 
The shape of the LMXB XLF is more complex. 
It appears to follow the $L^{-1}$ power law at low
luminosities, gradually steepens at $\log(L_X)\ge 37.0-37.5$ and has a
rather abrupt cut-off at $\log(L_X)\sim 39.0-39.5$. In the 
$\log(L_X)\sim 37.5-38.7$ luminosity range it has the differential
slope of $\approx 1.8-1.9$.  
This difference in the ``universal'' XLFs reflects
the difference in the accretion regimes  in  low- and
high-mass 
X-ray binaries. The majority of high mass systems are wind
accretors and their $\dot{M}$ distribution  is governed by
the properties of the optical companion, mainly by their
luminosity/mass distribution.\cite{postnov} 
The LMXBs, on the contrary, are close binaries fed via
Roche lobe overflow and the $\dot{M}$ in such systems is
generally defined by the angular momentum loss rate, which, in turn,
depends on the binary system parameters.

\subsection{Universal XLFs and binary evolution}

Existence of  ``universal'' XLFs is somewhat surprising. 
The shape of the luminosity function is defined by a number
of factor, such as metallicity and star formation history of the host
galaxy. The importance of the latter is illustrated by the 
example of low mass X-ray binaries. Indeed, a luminosity 
of $10^{38}$ erg/s requires a mass accretion 
rate of $\sim 10^{-8}$ M$_{\odot}$/yr, which can be sustained by a low
mass star for less than $\le 10^8$ yrs \cite{pods2002}.  This 
is significantly shorter than the life time of a galaxy. It is yet
shorter for the most luminous LMXB systems with $L_X\sim10^{39}$ 
erg/s. In order to have the presently observed shape of
the LMXB XLF, with a moderate fraction of sources with
$L_X\ge 10^{38}$ erg/s, in the stellar systems of the age of 
$\sim 10^{10}$ yrs, a continuous replenishment of the high luminosity
sources is required. Such a replenishment can be maintained, for
example, due to binary systems with initially less massive companion
stars, reaching the X-ray active phase at later times.  
An evolution of the luminosity function with time passed after the
star formation event must be present and it should be more pronounced
at the high luminosity end of the XLF. 
However, the results of Grimm et al.\cite{grimm03} and
Gilfanov\cite{lmxb}  suggest that there are no
significant galaxy-to galaxy variations of the shape of the luminosity
distributions of X-ray binaries in  nearby galaxies. 
A possible explanation is that the  nearby  galaxies
studied in Ref.~\citen{grimm03,lmxb} have similar ages of the stellar
populations, with the more subtle variations being masked by
statistical errors. The latter limitation is unavoidable
due to limited number of (bright) sources per galaxy.

\section{\boldmath $L_X-$SFR and $L_X-{\rm M}_*$ relations}

\begin{figure}
\centerline{\hbox{
\resizebox{0.5\hsize}{!}{\includegraphics{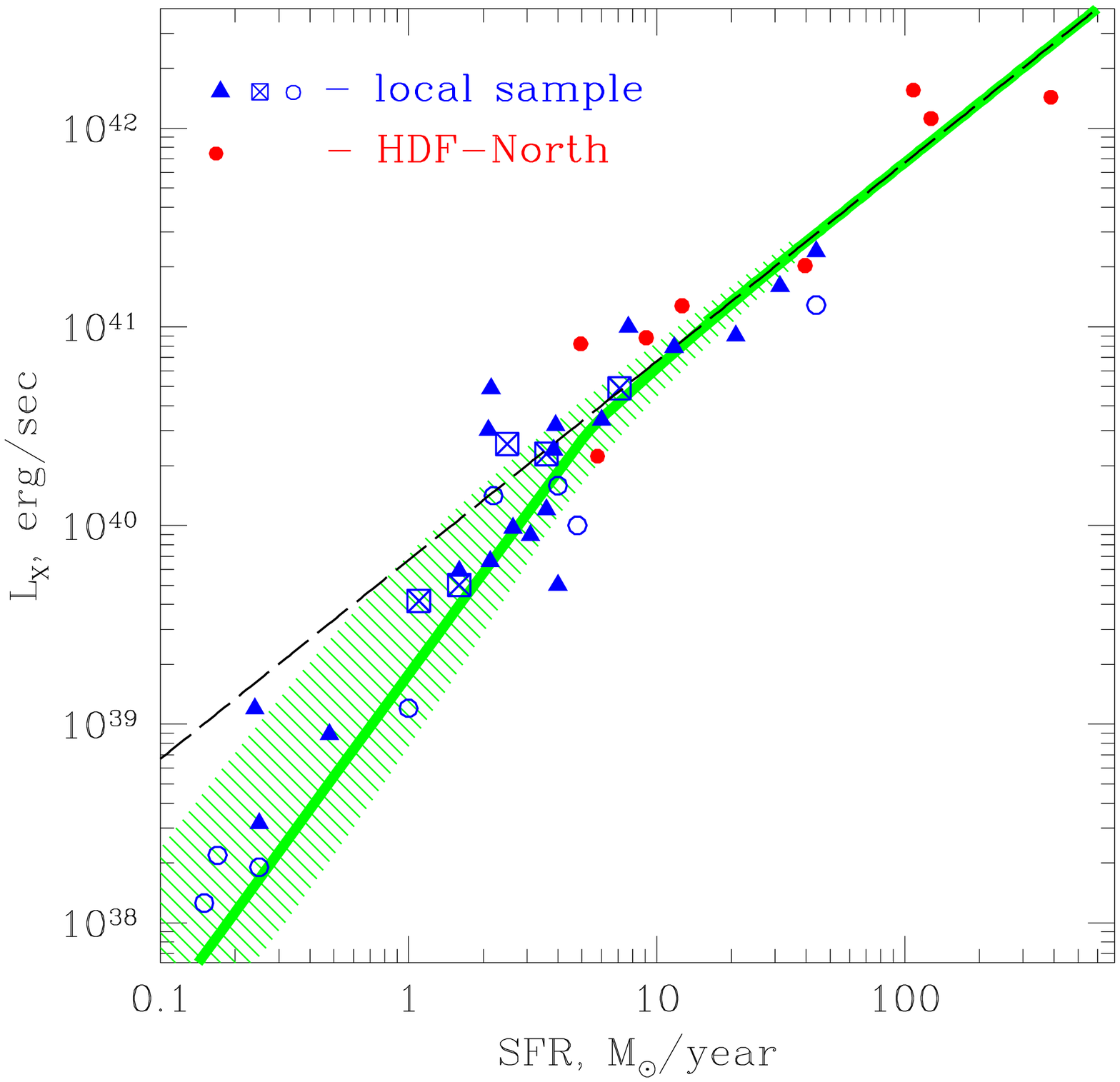}}
\resizebox{0.5\hsize}{!}{\includegraphics{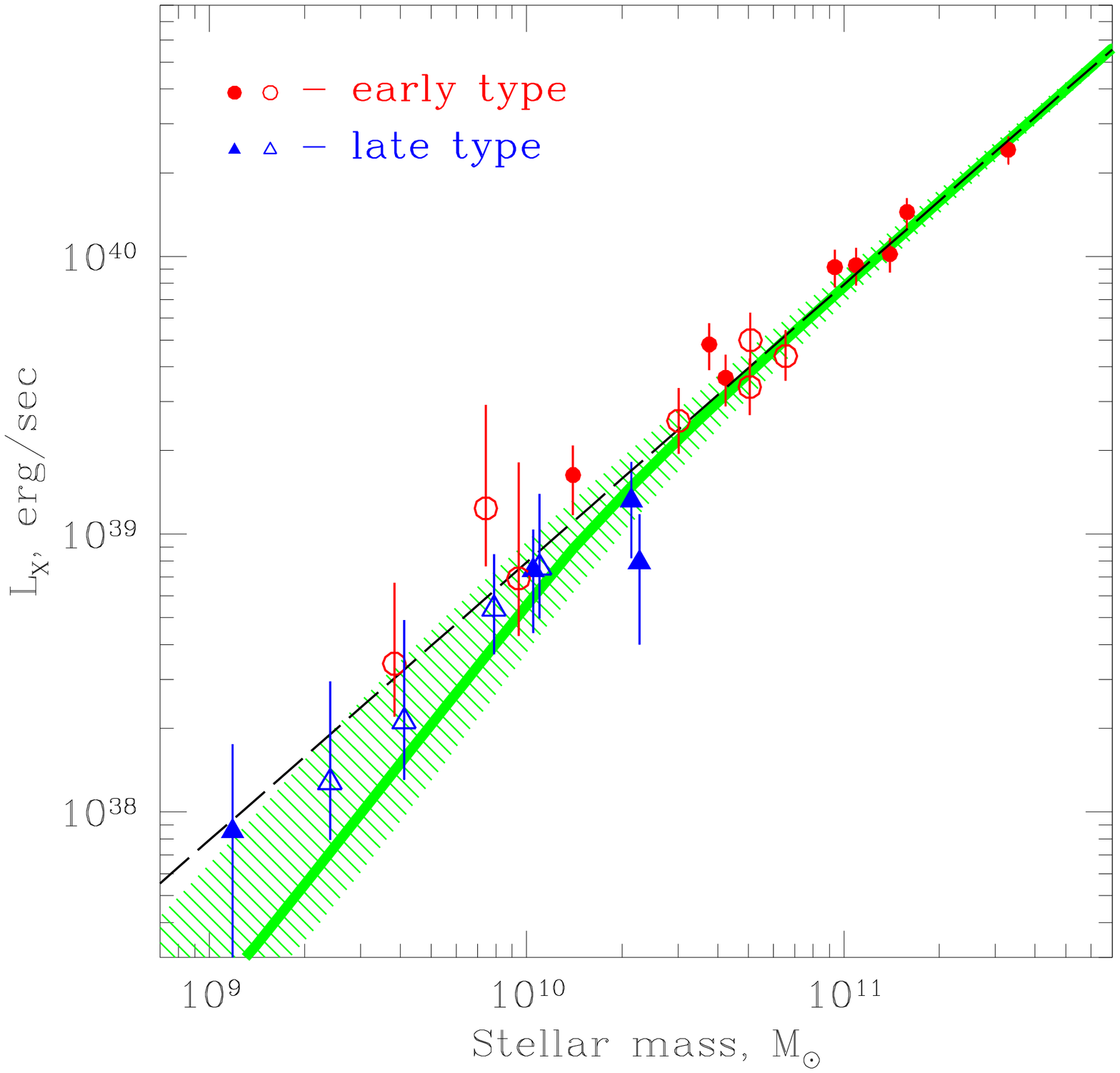}}
}}
\caption{The $L_X-$SFR ({\em left}) and $L_X-M_*$ ({\em right})
relations for high- and low-mass X-ray binaries respectively.
The thick grey lines and shaded areas show the relations for the most
probable value of the total luminosity and its 67\% intrinsic spread,
predicted from the respective ``universal'' XLFs,  
the dashed  lines show linear relations for the expectation
mean. From Grimm et al.\cite{grimm03} and Gilfanov\cite{lmxb}.
} 
\label{fig:lx-relations}
\end{figure}

The total numbers of LMXBs and HMXBs in a galaxy are proportional to its
stellar mass and star formation rate respectively. The same is 
true for the combined X-ray luminosities of X-ray binaries in the
limit of large SFR and stellar mass (Fig.~\ref{fig:lx-relations}). 
In the low SFR and stellar mass
regime the  $L_X-$SFR and $L_X-{\rm M}_*$ relations are modified by
the effects of small number statistics, as discussed below. The most
pronounced these effects are for HMXBs.

\begin{figure}
\centerline{\hbox{
\resizebox{0.5\hsize}{!}{\includegraphics{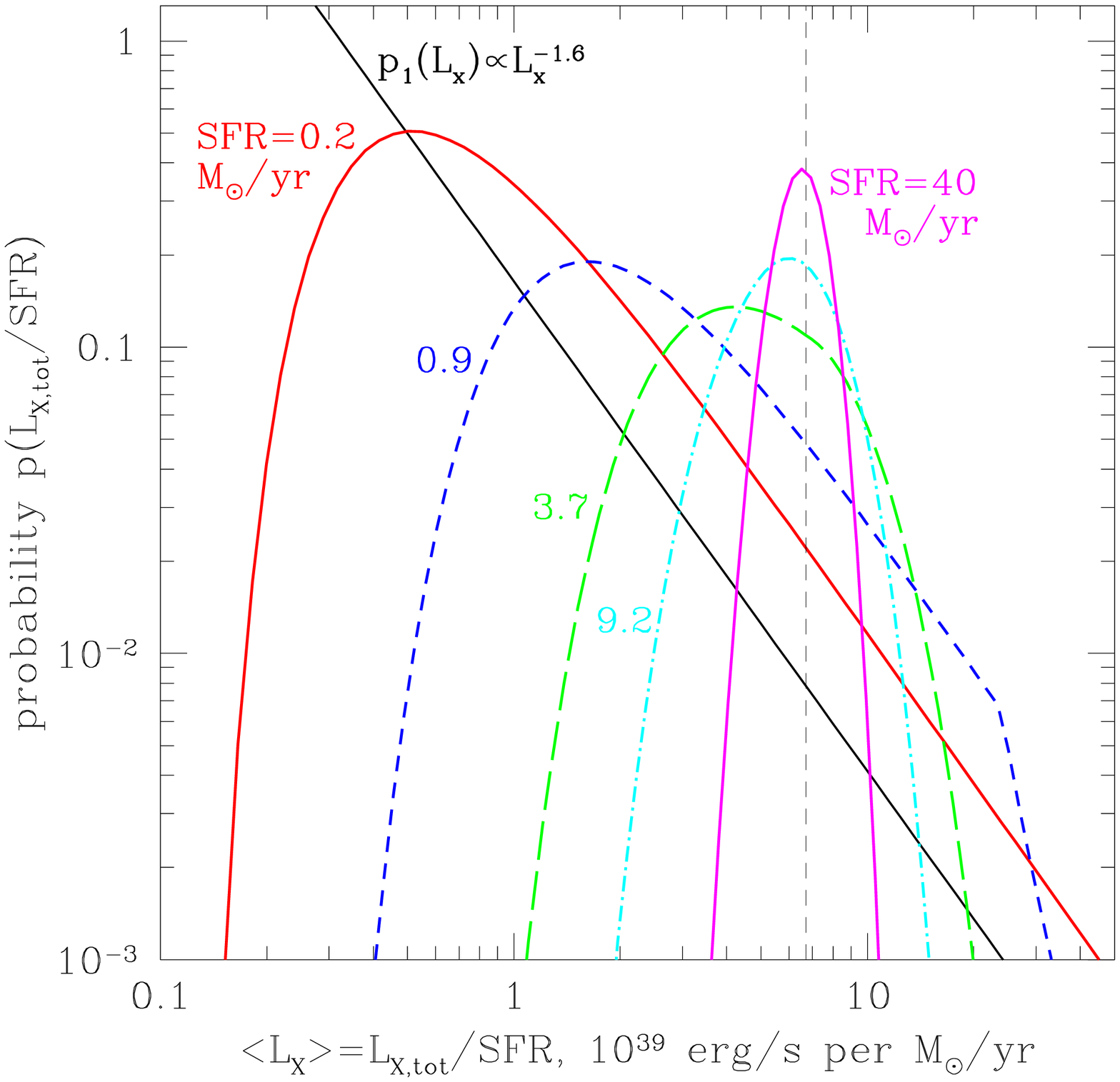}}
\resizebox{0.5\hsize}{!}{\includegraphics{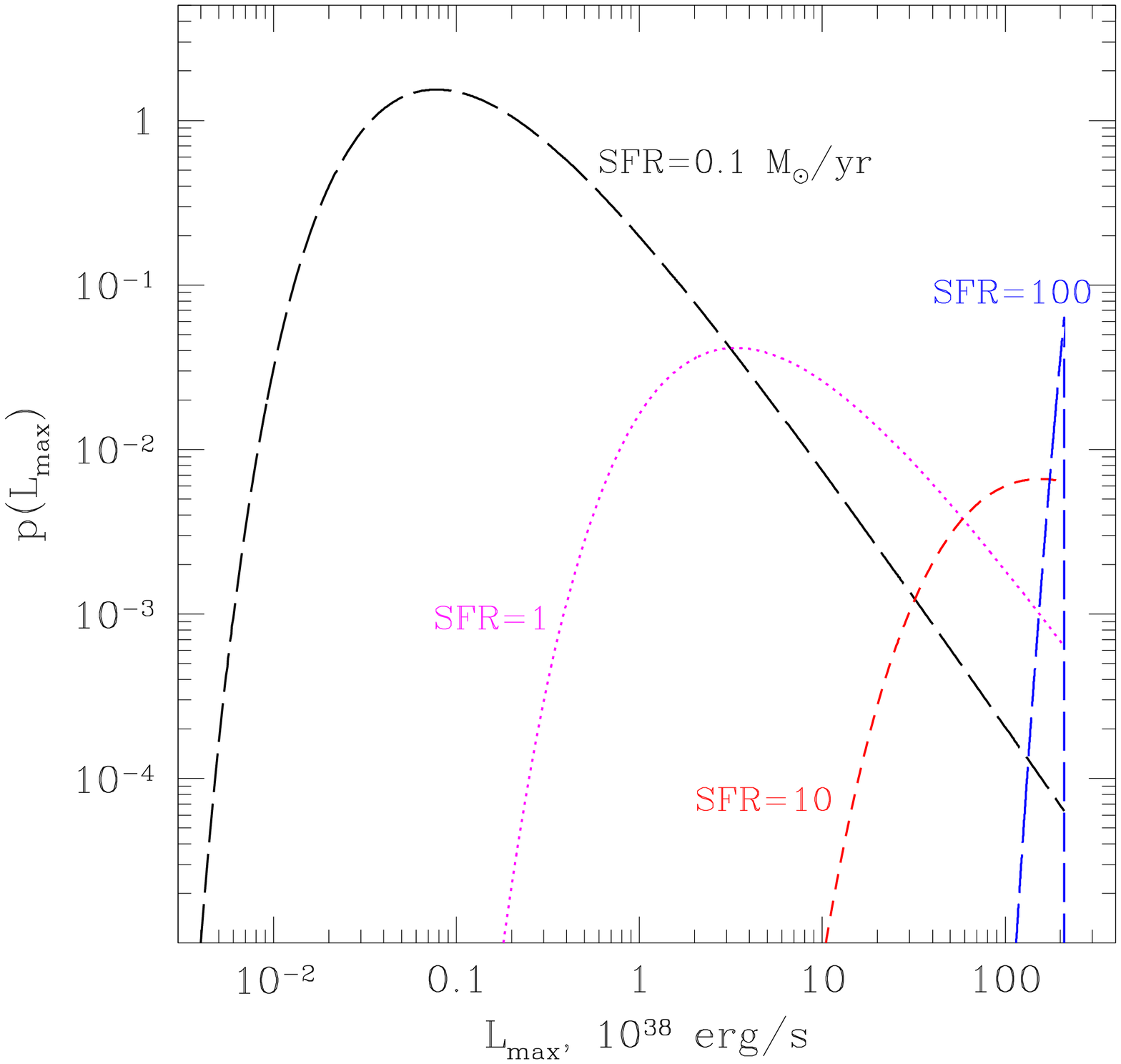}}
}}
\caption{
The probability distributions of the total luminosity of HMXBs
({\em Left}) and of the luminosity of
the brightest HMXB in a galaxy ({\em Right}) for different values of
SFR.  The vertical dashed line in the left panel shows the expectation
mean. Both distributions are computed using the parameters of the 
``universal'' luminosity function of HMXBs.  From Gilfanov et
al.\cite{stat}.   
} 
\label{fig:probdist}
\end{figure}

\subsection{Effects of small number statistics in the  $L_X-$SFR and
$L_X-M_*$ relations}

An intuitively obvious expression for the total luminosity can be
obtained integrating the luminosity distribution:
\begin{eqnarray}
\left  <L_{\rm tot} \right > =
\int_{L_{\rm min}}^{L_{\rm cut}} \frac{dN}{dL}\,L\,dL
\propto N_{\rm tot}
~~\propto ~~SFR {\rm ~ or} ~ M_*
\label{eq:ltot_mean}
\end{eqnarray}
implying, that the total luminosity is proportional to the 
number of sources, i.e. to the star formation rate or stellar
mass. However, as discussed by Gilfanov et al.\cite{stat}, the
quantity of interest  is a sum of  luminosities of discrete
sources: 
\begin{eqnarray}
L_{\rm tot}=\sum_{k} L_k
\label{eq:ltot_sum}
\end{eqnarray}
where $L_k$ are distributed according to the luminosity function
$dN/dL$. Depending on the properties of  $dN/dL$, the
probability distribution for the total luminosity, $p\,(L_{\rm tot})$,
can have a complex  shape. Most importantly, it
depends on the total number of sources and can be significantly
asymmetric for small SFR or 
${\rm M}_*$, as illustrated by
left panel in Fig.~\ref{fig:probdist}.  
Obviously,  the luminosity of a randomly chosen galaxy will most
likely be close to the value, at which $p\,(L_{\rm tot})$ has the
maximum (the mode of the probability distribution). 
Owing to the skewness of the  $p\,(L_{\rm tot})$ in the low-SFR and
low-$\rm M_*$ regime, its mode does not equal the expectation
mean defined by eq.~\ref{eq:ltot_mean}. 
If observations of many  galaxies with the same star formation rate
(or stellar mass)  are performed, the  measured values of $L_{\rm tot}$
will be distributed according to $p\,(L_{\rm tot})$. Their average
will be always equal to the expectation
mean,  shown by the dashed straight
line in the left  panel of Fig.\ref{fig:probdist}. 
Of course, in the limit of
$N_{\rm tot}\rightarrow\infty$ the $p\,(L_{\rm tot})$ asymptotically
approaches the normal distribution, in accord with the Central Limit
Theorem.  

The difference between the mode and expectation mean is further
illustrated by Fig.\ref{fig:lx-relations}, showing the predicted
$L_X-$SFR  and $L_X-{\rm M_*}$ relations along with the data of Chandra
observations  of nearby and (for HMXBs) distant galaxies in HDF-N. 
The thick solid lines in  the figure  show the SFR and $\rm M_*$
dependence of the  mode of the $p\,(L_{\rm   tot})$ and 
predict the {\em most probable} value of the X-ray  luminosity of a
{\em randomly chosen} galaxy. The dashed lines show the expectation mean --
the average luminosity of a large number of galaxies with the same SFR
or $\rm M_*$, which depend linearly on the SFR and $\rm M_*$.
For the shape of the LMXB XLF, the total luminosity of LMXBs is
defined by the sources with luminosity $\log(L_X)\sim
37-37.5$. Correspondingly, the effects of small number statistics,
although present in the $L_X-M_*$ relation,  are less important than
in the $L_X-$SFR relation (Fig.~\ref{fig:lx-relations}). 

Validity of the $L_X-$SFR relation beyond the local Universe was
confirmed by the Chandra observations of star forming galaxies at
$z\approx 0.2-1.3$ in HDF-N. The $L_X-M_*$ relation for LMXBs was
studied for local galaxies only, within $\sim 20-30$ Mpc from the
Sun. Due to binary evolution effects,  this relation can be different
for younger galaxies, located at intermediate redshifts.

\begin{figure}
\centerline{
\resizebox{0.5\hsize}{!}{\includegraphics{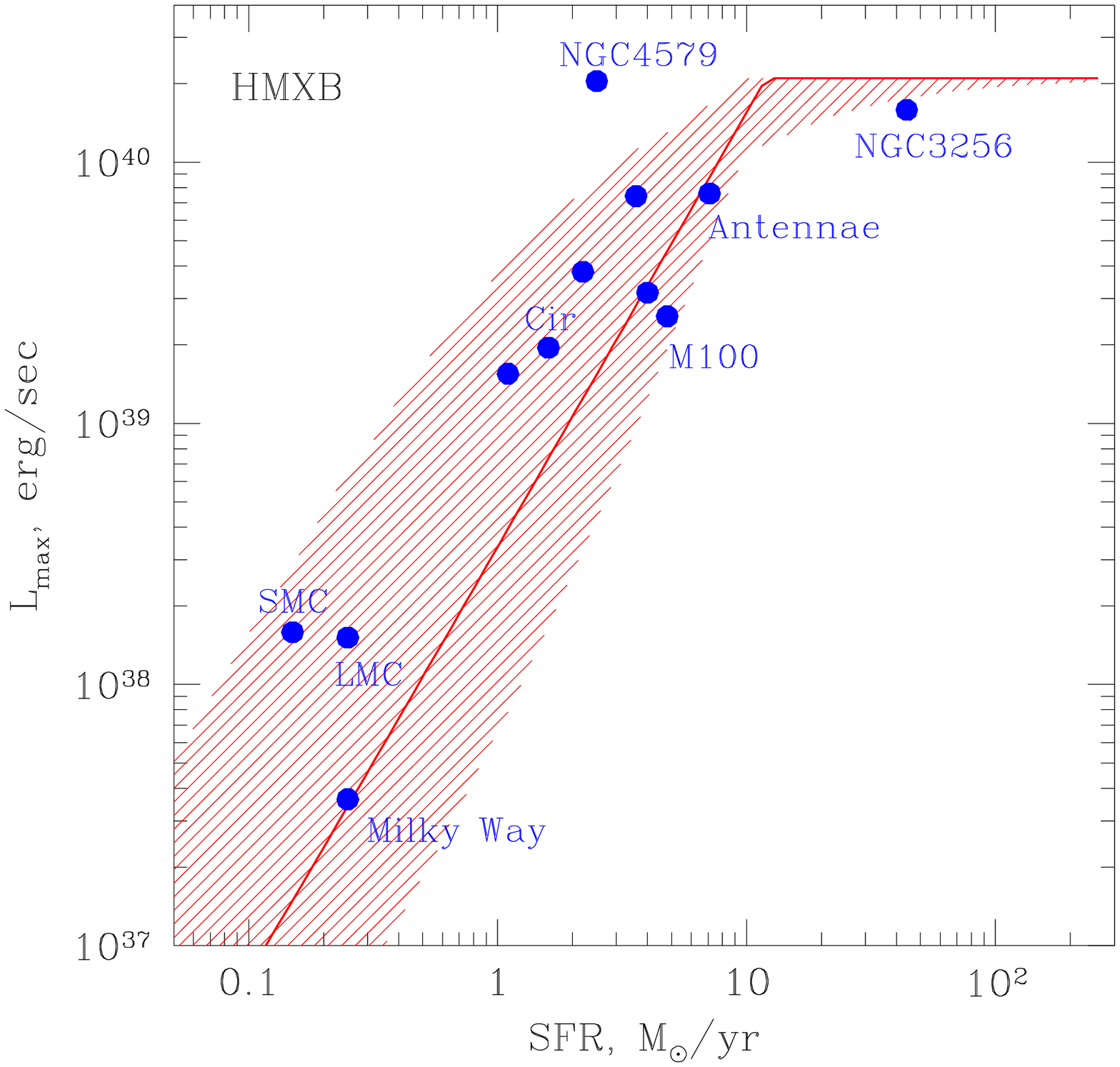}}
\resizebox{0.5\hsize}{!}{\includegraphics{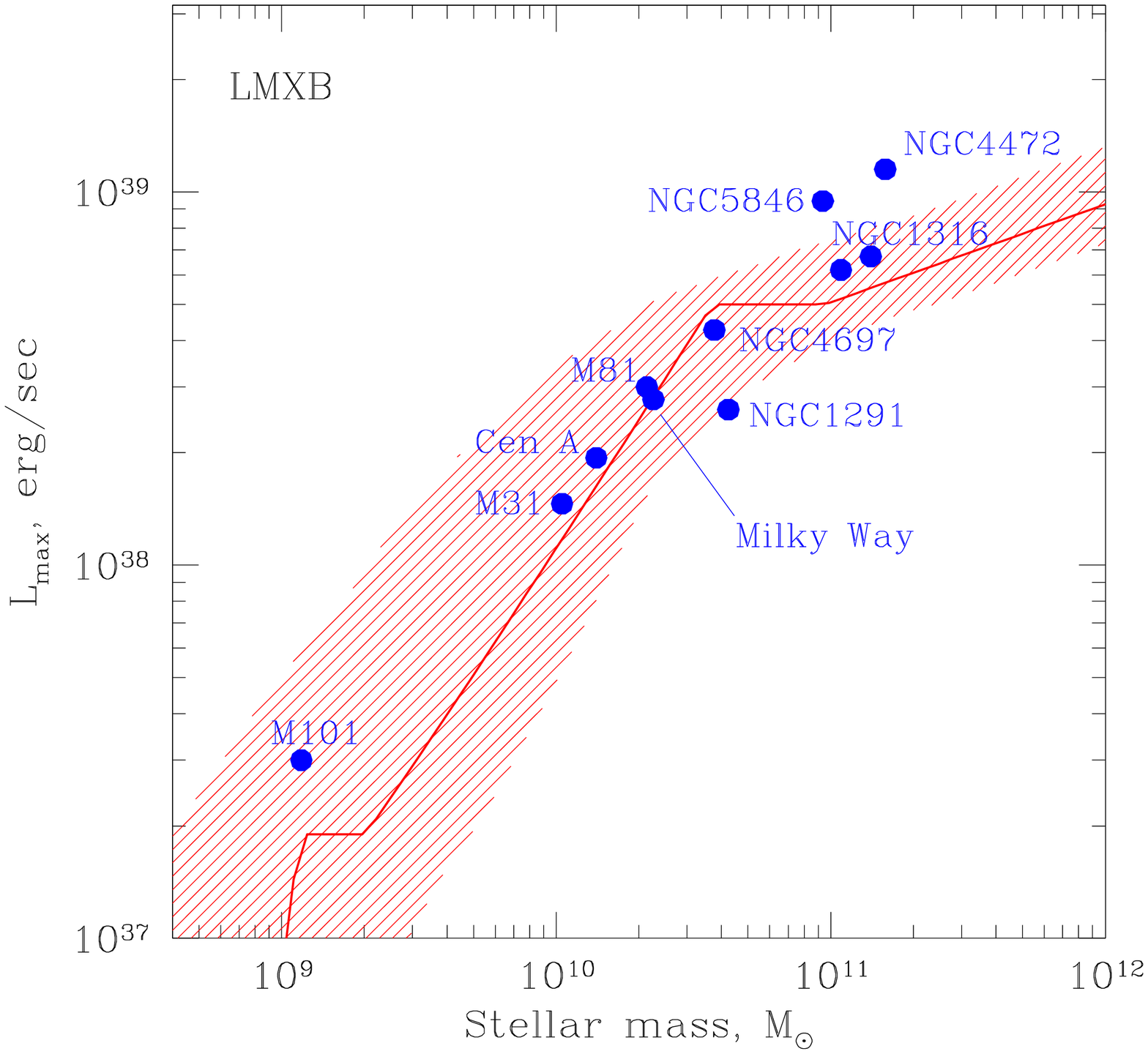}}
}
\caption{{\em Left:} 
The luminosity of the brightest HMXB ({\em
left}) and LMXB ({\em right}) in a galaxy as a function of the star
formation rate and the stellar mass respectively.
The solid lines and the shaded areas show the most probable value 
of the $L_{\rm max}$ and its 67\% intrinsic dispersion, calculated
from the respective ``universal'' XLFs.
The filled circles show $L_{\rm max}$ observed in the nearby galaxies
and in the Milky Way. 
The ``broken'' shape of the predicted dependence in the right panel is
a consequence of the broken power law approximation of the
``universal'' LMXB XLF, used in the calculations.
From Gilfanov\cite{stat}. 
}
\label{fig:lmax}
\end{figure}

\section{Luminosity of the brightest source in a galaxy}

As the first Chandra observations of compact sources in nearby
galaxies became available, it has been noted 
\cite{ngc4697,ngc1291,fabbiano2003} that the
luminosity of the brightest  X-ray binary in a galaxy might depend on
its properties. It appeared to correlate with the star formation rate
(HMXBs) and stellar mass (LMXBs) of the host galaxy. For example, in
the Antennae galaxies, a number of compact sources have been
discovered with luminosities of $\sim 10^{40}$ 
erg/s \cite{ant}. On the other hand, the luminosities of the
brightest HMXB sources in the Milky Way do not exceed $\le 10^{38}$
erg/s \cite{grimm02}. It has been argued that this might reflect the
difference in the intrinsic source properties, related to the
SFR-dependent difference in the galactic environment and in initial
conditions for X-ray binary formation. 

However, as was noted in Ref.~\citen{lmxb}, the probability
distribution for the luminosity of the brightest source in a galaxy,   
$p\,(L_{\rm max})$, depends non-trivially on the LF normalization,
i.e. on the SFR or the stellar mass of the host galaxy (right panel in
Fig.~\ref{fig:probdist}).
This leads to the dependence of the most probable value of the
luminosity of the brightest source on, for example, the star formation
rate of the galaxy -- $L_{\rm max}$ increases with SFR, until it 
reaches the maximum possible value, defined by the high
luminosity cut-off of the LF, as illustrated by Fig.~\ref{fig:lmax}.
Filled symbols in Fig.\ref{fig:lmax} show the luminosities of the 
brightest source observed with Chandra in nearby galaxies.
The large difference in the maximum luminosity  between low-
and high-SFR galaxies, e.g. between the Milky Way and the Antennae
galaxies, or between massive elliptical galaxies and the bulges of
spiral galaxies can be naturally understood in terms of the properties
of the probability distribution $p\,(L_{\rm max})$.  
So far there is no evidence for the significant dependence of
intrinsic properties of X-ray binaries on the galactic environment.

\section{Variability of the total emission of X-ray binaries}

X-ray flux from X-ray binaries is known to be variable in a broad
range of time scales, from $\sim$msec to $\sim$ years.
In addition to a number of coherent phenomena and 
quasi-periodic oscillations, significant continuum  
aperiodic variability is often observed. The fractional rms of
aperiodic variations depends on the nature of the binary system and
the spectral state of the X-ray source and is usually in the range
from  a fraction of a per cent to $\sim 20-30$ per
cent. 

\begin{wrapfigure}{l}{6.6cm}
\centerline{
\resizebox{\hsize}{!}{\includegraphics{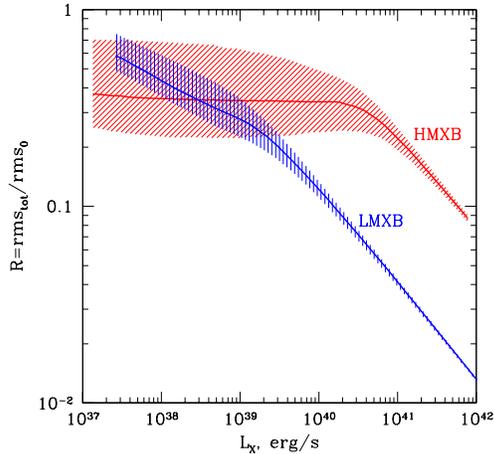}}
}
\caption{Fractional $rms$  of combined emission of HMXBs
and LMXBs as a function of their total luminosity. 
The $rms_0$ denotes  $rms$ of one source, assumed to be the same for
all sources.  
The thick solid lines and shaded areas show the most probable value of
the $rms/rms_0$, and its 67\%  confidence interval, both obtained from
the Monte-Carlo simulations for the respective ``universal''
XLFs.  At high $L_X$ both curves follow
$rms\propto 1/\sqrt{L_X}$ averaging law. 
}
\label{fig:rms_ltot}
\end{wrapfigure}

As flux variations of the individual sources are uncorrelated, one
might expect that the fractional $rms$ of the total emission should
decrease with the number of sources $n$ as 
$rms\propto 1/\sqrt{n}\propto 1/\sqrt{L_{\rm tot}}$.
Although correct in the limit of large $n$, this intuitively obvious
prediction can break down for small number of sources.
Indeed, for a sufficiently flat luminosity function, a regime exists,
when the total luminosity is defined by a few brightest sources. 
To first approximation the number of such sources does not depend on
the total number of sources.
Consequently, in this regime the fractional $rms$ of the total
emission depends weakly or  does not depend at all
on the total number of sources or on their total luminosity (i.e. on
SFR or $M_*$). 
This behavior is illustrated by the results of the Monte-Carlo
simulations, shown in Fig.\ref{fig:rms_ltot} as $rms-L_{\rm tot}$
relations for HMXBs and LMXBs. 

For moderate star formation rates, 
SFR$\le 5-10~M_{\odot}$/yr, we predict a rather large aperiodic
variability of the total emission of HMXBs at the level of $\sim
1/3-1/2$ of the fractional $rms$ of individual X-ray binaries. 
At larger values of SFR, corresponding to the linear regime in the
$L_X-$SFR relation, it decreases as 
$rms\propto 1/\sqrt{\rm SFR}\propto 1/\sqrt{L_{\rm tot}}$, 
in accord with the averaging law. 
For LMXBs, owing to the shape of their ``universal'' XLF, the
fractional $rms$ of the total emission decrease rather quickly with
$M_*$ and  $L_{\rm tot}$  in the entire mass range of interest.
Consequently, considerable variability on the level of $\sim 1/4-1/2$
of that of individual X-ray binaries can be expected only for light
bulges of spiral galaxies with masses in the $\log(M_*)\sim 9.5-10.5$
range. 
In the bright luminosity end, $\log(L_X)\ge
39.5$, the X-ray emission from early type galaxies is expected to be
significantly, up to a factor of $\sim 7$, less variable than from
star forming galaxies.
The predicted $rms-L_X$ relations can be
modified by the luminosity dependence of the  $rms$ of individual
sources. This factor might become especially important for HMXBs at
large values of SFR when the total luminosity of a star forming galaxy
is dominated by ULXs whose variability properties we know little
about.

\begin{figure}[t]
\centerline{\hbox{
\resizebox{0.45\hsize}{!}{\includegraphics{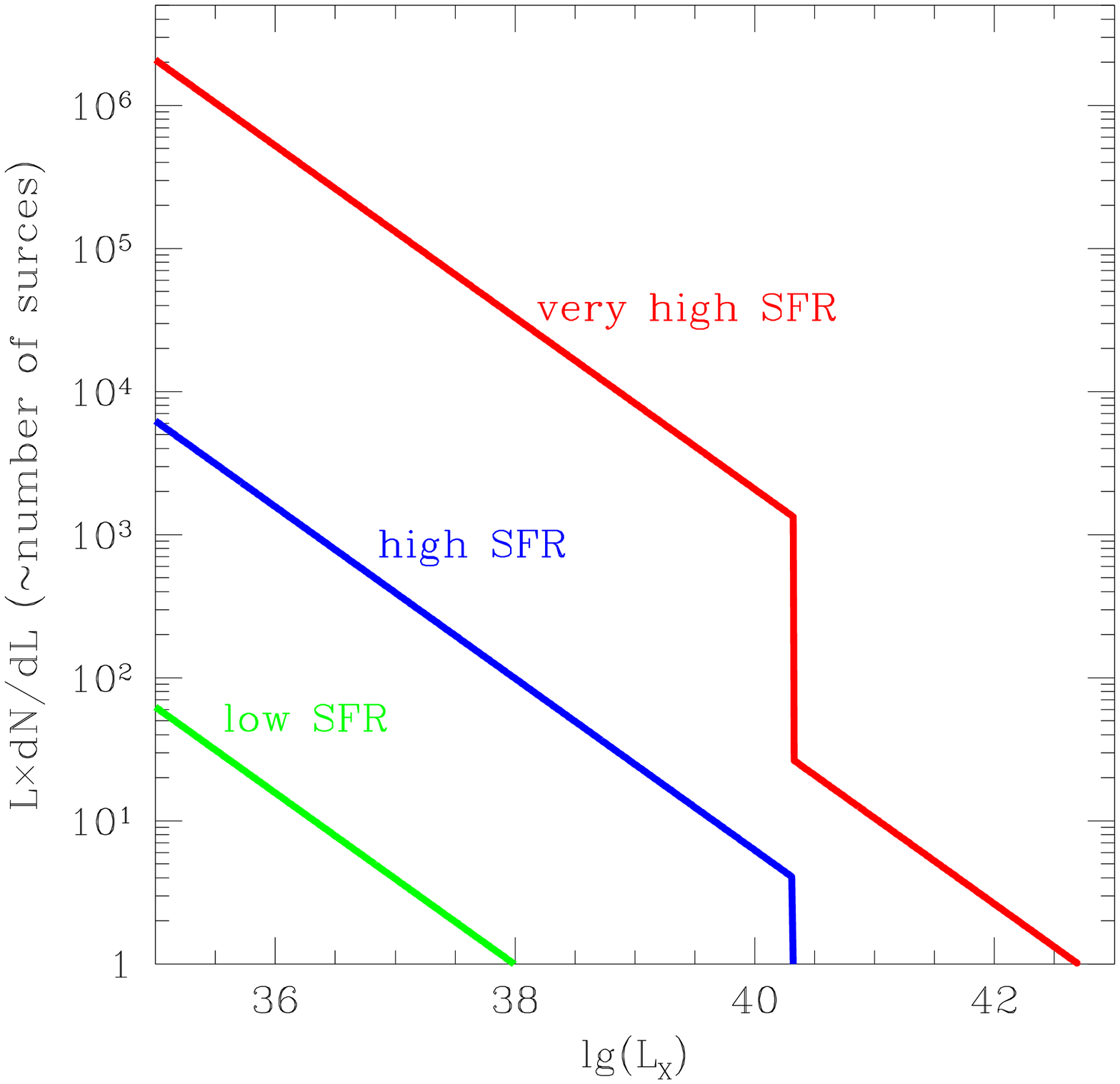}}
\resizebox{0.45\hsize}{!}{\includegraphics{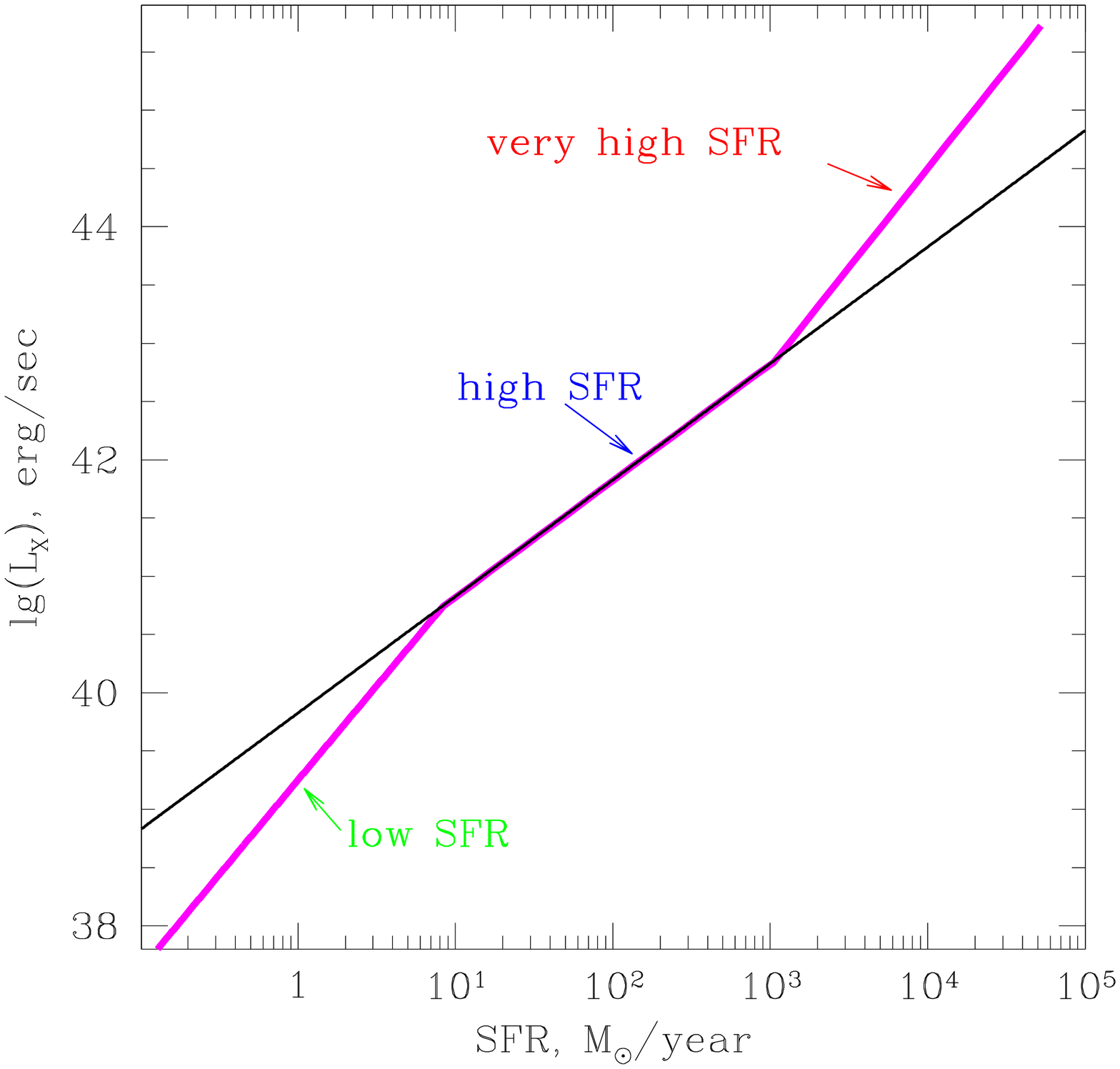}}
}}
\caption{Illustration of the effect of hypothetical intermediate mass
black holes on the $L_X-$SFR relation. 
{\em Left:} The luminosity function of compact sources at different
levels of star formation rate. 
{\em Right:} Corresponding $L_X-$SFR relation. The thin straight line
shows the linear dependence.
\label{fig:imbh}}
\end{figure}

\section{X-ray binaries, ULXs and intermediate mass black holes}

The cut-off luminosity in the LMXB XLF is by a factor of $\sim 10$
smaller than in HMXB XLF (Fig.~\ref{fig:xlf_dif}) and does not exceed
$\log(L_X)\sim 39.5$.  Earlier report of detection of 
the sources with luminosity significantly greater than $\sim 10^{39}$
erg/s in early type galaxies are most likely related to
miss-identifications of the CXB sources, erroneously attributed to the
galactic source populations.\cite{lmxb,ulx_ell} 
Quantitatively, the value of the high luminosity cut-off in the LMXB
XLF, $\log(L_X)\sim 39-39.5$  does not present a problem from the
point of view of the Eddington luminosity limit for a stellar mass
object and can be easily explained by a sub- or near-critical
accretion onto a $\sim  10-15~M_\odot$ black hole --
there are no ``real'' ULXs in old stellar systems.
The truly exotic are the objects associated with the regions of
intensive star formation, having luminosities above $\sim 10^{40}$
erg/s.

Surprising is the smooth, single slope power law shape of the
``universal'' luminosity function of  X-ray sources in star forming
galaxies, without any significant steps and features in a broad
luminosity range, $\lg(L_X)\sim 36-40.5$ (Fig.\ref{fig:xlf_dif}).
The high luminosity end, $\lg(L_X)>39$, of this distribution 
corresponds to ultraluminous X-ray sources. 
Its low luminosity end, on the other hand, is composed of ordinary
X-ray binaries, powered by accretion onto a $\sim$stellar mass compact
objects.  
This result  constrains the range of  possible models for ULXs. 
Their frequency and luminosity distributions should be a smooth
extension toward higher luminosities of that of ``ordinary''
$\sim$stellar mass systems, emerging from the standard stellar
evolution sequence.   
Although some of the ULXs might be indeed rare and exotic objects,
it appears that majority of them cannot be a completely
different type of the source population, but, rather,  represent the
high mass, high $\dot{M}$ tail of the ``ordinary'' HMXB population.

The position of the break between the non-linear and linear parts of
the $L_X$--SFR relation depends on the LF slope and its cut-off
luminosity: ${\rm SFR}_{\rm break}\propto L_{\rm cut}^{\alpha-1}$. 
This allows one to constrain the parameters of the luminosity
distribution  
of compact sources using the data of spatially unresolved galaxies. 
Agreement of the predicted $L_X-$SFR relation with the data both
in high- and low-SFR regimes confirms the universality of the HMXB
luminosity function, derived by Grimm et al.\cite{grimm03} from
significantly fewer 
galaxies (shown as crossed boxes) than plotted in
Fig.\ref{fig:lx-relations}. It provides an
independent confirmation of the existence of a cut-off in the HMXB
XLF at $\log(L_{\rm cut})\sim 40.5$, thus confirming that the
luminosity of ULXs in the nearby galaxies has a maximum value of  
the order of $\lg(L_X)\sim 40.5$.
The fact that the (spatially unresolved) galaxies from the
Hubble Deep Field North obey the same $L_X-$SFR relation
(Fig.\ref{fig:lx-relations}), implies, that the ULXs at the redshift
of $z\sim 0.2-1.3$ were not significantly more luminous, that those
observed in the nearby galaxies.

The hypothetical intermediate mass black holes, probably reaching
masses of $\sim 10^{2-5} M_{\odot}$, might be produced, e.g. via black 
hole merges in dense stellar clusters, and can be associated with 
extremely high star formation rates. To accrete efficiently they
should form close binary systems with normal stars or be located  in
dense molecular clouds. It is natural to expect, that such objects are
significantly less frequent than $\sim$stellar mass black holes. 
The transition from $\sim$stellar mass BH HMXB to intermediate
mass BHs should manifest itself as a step in the luminosity
distribution of compact sources (Fig.\ref{fig:imbh}, left panel).  
If the cut-off in the HMXB XLF, observed at $\log(L_{\rm cut}) \sim
40.5 $ corresponds to the maximum possible luminosity of ``ordinary''
$\sim$stellar mass black holes and if at $L>L_{\rm cut}$ a population of 
hypothetical intermediate mass BHs emerges, it should lead to
a drastic change in the slope of the $L_X$--SFR relation at extreme
values of SFR  (Fig.\ref{fig:imbh}, right panel). 
Therefore, observations of $L_X-$SFR relation for distant star forming
galaxies with very high SFR might be a way to probe the
population of intermediate mass black holes.


%


\begin{thebibliography}{99}
  


\bibitem{colbert99} 
E.~J.~M.~Colbert \& R.~F.~Mushotzky, \JL{ApJ,519,1999,89}

\bibitem{colbert04} 
E.~J.~M.~Colbert et al. \JL{ApJ,602,2004,231}


\bibitem{fab89} 
G.~Fabbiano, \JL{ARA\&A,27,1989,87}

\bibitem{fabbiano2003} 
G.~Fabbiano \& N.~White, (2003), in: ``Compact Stellar
X-ray Sources'', Cambridge University Press (eds., W. Lewin \& M. van
der Klis), astro-ph/0307077

\bibitem{lx-sfr} 
M.~Gilfanov, H.-J.~Grimm  \& R.~Sunyaev, \JL{MNRAS,347,2004,L57}

\bibitem{lmxb} 
M.~Gilfanov, \JL{MNRAS,349,2004,146}

\bibitem{stat} 
M.~Gilfanov, H.-J.~Grimm  \& R.~Sunyaev, MNRAS (2004), submitted,
astro-ph/0312540 

\bibitem{grimm02} 
H.-J.~Grimm, M.~Gilfanov \&  R.~Sunyaev, \JL{A\&A,391,2002,923}

\bibitem{grimm03} 
H.-J.~Grimm, M.~Gilfanov \& R.~Sunyaev, \JL{MNRAS,339,2003,793}

\bibitem{ngc1291} 
J.~A.~Irwin, C.~L.~Sarazin  \& J.~N.~Bregman, \JL{ApJ,570,2003,152}

\bibitem{ulx_ell} 
J.~A.~Irwin, J.~N.~Bregman  \& A.~E.~Athey, \JL{ApJ,601,2004,L143}

\bibitem{kalogera2004} 
V.~Kalogera et al. \JL{ApJ,603,2004,41}

\bibitem{kim2004} 
D.-W.~Kim \& G.~Fabbiano, ApJ (2004), submitted, astro-ph/0312104

\bibitem{king2001} 
A.~R.~King et al. \JL{ApJ,552,2001,109}

\bibitem{koerding2002} 
E.~Koerding, H.~Falcke, S.~Markoff \JL{A\&A,382,2002,L13}

\bibitem{miller2003} 
J.~M.~Miller et al. \JL{ApJ,585,2003,37}

\bibitem{pods2002} 
Ph.~Podsiadlowski, S.~Rappaport  \& E.~D.~Pfahl, \JL{ApJ,565,2002,1107}

\bibitem{postnov} 
K.~Postnov, \JL{Ast.Lett.,29,2003,372}

\bibitem{ranalli} 
P.~Ranalli, A.~Comastri  \& G.~Seti, \JL{A\&A,399,2003,39}

\bibitem{ngc4697} 
C.~L.~Sarazin, J.~A.~Irwin \& J.~N.~Bregman,  \JL{ApJ,556,2001,533} 

\bibitem{rs78} 
R.~A.~Sunyaev, B.~M.~Tinsley  and D.~M.~Meier, 
\JL{Comment.~Astrophys.,7,1978,183}

\bibitem{xrbrev} 
F.~Verbunt \& E.~P.~J~van den Heuvel, (1995), in:
X-ray Binaries, Eds: W.Lewin, J.van Paradijs \& E.van den Heuvel,
Canbridge Univ.Press, p.457

\bibitem{ant}
A.~Zesas et al., \JL{ApJS,142,2002,239}


\end{thebibliography}
\end{document}